# Magnetic Shield Approach Toward Non-adiabatic Low to High Field Positron Beam Transition for Microtraps

M. A. Khamehchi[1], A. Narimannezhad, M. H. Weber, and K. G. Lynn
Center for Materials Research, Washington State University, 99164-2711, USA

## Abstract

This simulation study sheds light on effectiveness of a magnetic shield on non-adiabatic transport of a positron beam from $3700G$ to $100G$ where we were able to maximize the low to high field transmission efficiency for $6.5KeV$ positrons. An initially brighter beam can be transported even more efficiently. Cross-sectional uniformity of the beam is also studied for different beam energies. This is important for filling microtraps consist of thousands tubes with diameters of tens of microns each. Transverse momentum of the particles is also studied since it causes mirroring in the high field and compromises efficiency. Beam remoderation is required as $\sim eV$ energy positrons will be trapped in the Microtrap. The observed magnetic shield assisted low to high filed transition opens new avenue for high efficiency manipulating of non-neutral beams.

## 1 Introduction

### 1.1 Statement of Problem

New energy visions have been intensively sought after in the last few decades. In fact, new advanced, clean and sustainable technologies have been the peak of interest over the last several years. No energy source to date can compare to antimatter annihilations, a reaction that is 100% efficient. Positrons (anti-electrons) are of great interest due to the readily available and easy to produce electron. Penning-Malmberg (PM) traps [1][2] are widely used for storing non-neutral plasmas such as positron plasmas because of the ease of construction using a static magnetic field. A schematic sketch of a PM trap biased to a potential of $\phi_0$ and placed in the constant magnetic field of magnitude $B$, is shown in Figure 1.

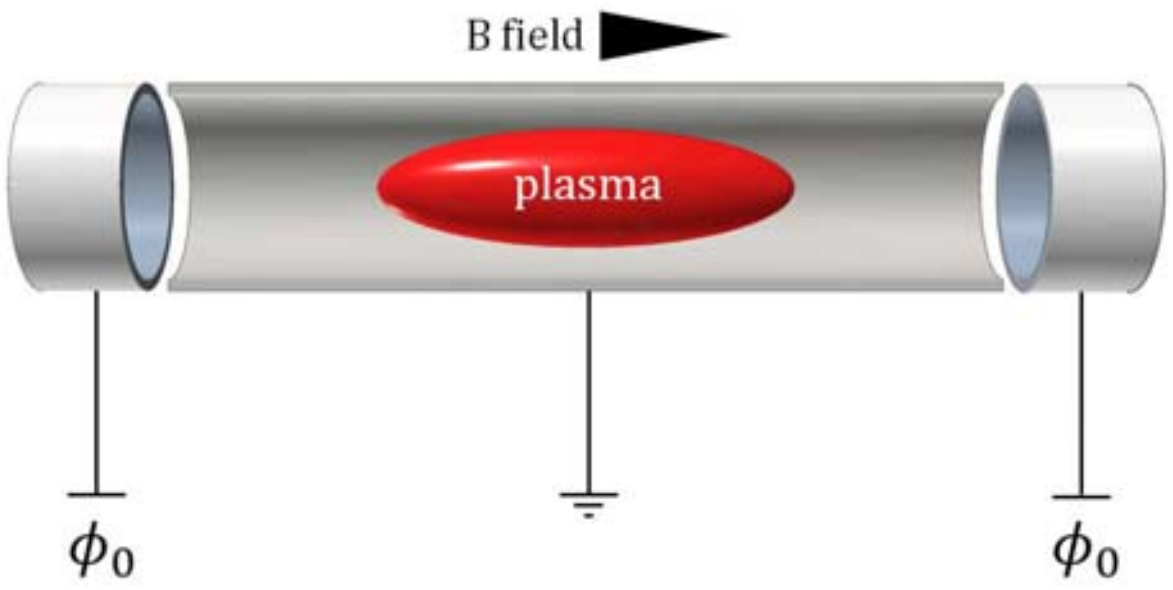

**Figure 1 A schematic sketch of a Penning-Malmberg trap with $\phi_0$ as the end cap potential in the constant magnetic field of B.**

[1] mohammad.khamehchi@wsu.edu

During recent years low energy positron sources are becoming increasingly important in many areas of physics and technology, including atomic physics research [3], plasma physics [4], astrophysics [5], mass spectrometry [6], antihydrogen physics [7], and materials research [8][9]. In many cases tunable positron sources, both in energy and intensity, are necessary and can be achieved by using PM traps [10]. Unavoidable limitations on the number of particles in a single trap device has led to a new design of proposed by one of the authors (K. G. Lynn) which consists of thousands of parallel microtubes placed in a closed packed format, and being filled in the beam line [11][12][13]. The microtubes have large length to radius aspect ratio O(1000:1) and the whole trap poses a low confinement voltage O(10 V).

Confining potential ($\phi_0$) at the two end electrodes of a PM trap can be obtained from

$$\phi_0 = \frac{eN_p}{4\pi\varepsilon_0 L_p}\left[2ln\left(\frac{a}{\rho_p}\right)+1\right] \tag{1}$$

where $e$ is the positron charge, $N_p$ is the number of positrons, $L_p$ is the plasma length, $\varepsilon_0$ is the vacuum permittivity, and $a$ and $\rho_p$ are the radius of the trap and the plasma [11]. Thus, for $10^{13}$ positrons in a trap of length $10cm$ and even with $(a/\rho_p) = 1$ which is practically impossible, the confining potential is $1.4 \times 10^5 V$ which is hard to maintain especially for a portable device. Using micro-traps reduces this amount significantly since the trap walls tend to neutralize the space charge. Recently, simulations have shown the feasibility of making such devices with enhanced total number of particles with a confining potential of a few tens of volts [14] [15]. Figure 2 shows the concept of a compact positron trap. The long aspect ratio micro-traps are created via stacking up identical etched conductive wafers [16]. The wafers are surrounded by a radiation shield, and a superconductor magnet to provide the static solenoid magnetic field. The design of this modified, short PM trap with very small space charge potentials reduces plasma heating and so improves plasma confinement.

Recent trapping results in our research group agree with the simulations. Experimental results will be published in a future paper(s). Micro-Penning-Malmberg traps are truly an advanced method for storing much higher densities of antimatter. If we ever want to take advantage of the greatest energy dense source known to man, we must find ways to increase the density of antimatter plasmas. Higher densities with lower confining potentials is the key to making a transportable storage device which can propel this technology into the future. Clean energy, space exploration, advanced material interrogations, antimatter propulsion and controllable reactions are all going to benefit from this technology.

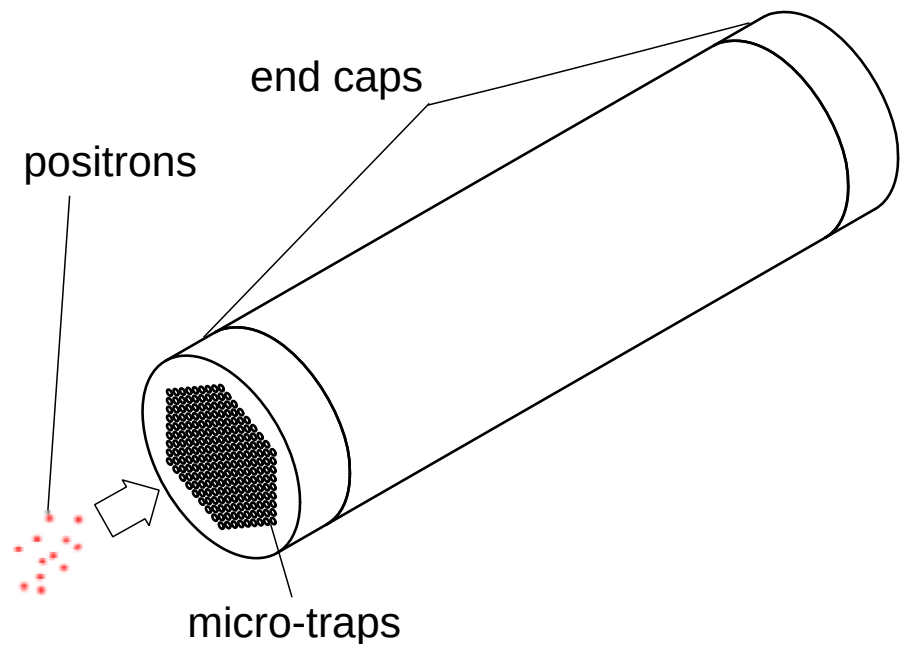

**Figure 2 The microtrap concept. Drilled conductor wafers are stacked up to build thousands of parallel long aspect ratio micro-traps.**

The use of high magnetic fields in these PM traps reduces positron annihilation on the trap walls and therefore insures longer trapping lifetimes. Also increasing the magnetic field increases the Brillouin limit in the plasma and more particles can be stored [14].

## 1.2 Beam Divergence

Higher magnetic field has a few disadvantages including more particle mirroring and more beam contraction. Mirroring reduces transport efficiency while contracting reduces the trapping efficiency since only a small portion of the micro-tubes will be filled. To fill more traps in parallel, a larger beam diameter is favorable. This also eliminates the complications with beam rastering or dioctron excitation.

There are several methods to “diverge” a particle beam when transitioning to extremely high fields and injecting them into the holes. In all methods if the beam is centered to the holes or if it has a larger diameter than one hole (filling multiple holes at once), the plasma does not experience diocotron modes which has deteriorative effect of the particles lifetime.

First method is beam rastering. Using ExB plates, we can raster the beam quickly to address specific holes and fill multiple tubes at once. Simulation has been done using the SIMION® simulator and ExB plates are already designed and fabricated. But it is hard to raster the beam exactly at the center of each microtube. To improve the alignment, MEMS technology can be employed on SiC to form dimples that resemble a muffin tin. This can be done such that the dimples align with the microtubes. The SiC disk would be installed directly in front of the microtrap. As positrons hit the SiC, they are channeled to the microtubes. As they drop out the other side, they are aligned to the microtubes.

Second and more suitable solution is using a Mu metal spider. The field terminates while allowing the majority of the particles to pass through without shrinking the beam too much. Several holes are filled at the same time with a larger diameter beam and so the diocotron modes are not present. Mu metal field termination also prevents particles mirroring as entering to high magnetic fields. High efficiencies are possible with careful control and manipulation of the transfer process, which is presented by simulation studies in this paper.

Researches were also done on delivering the plasma to the off-axis traps in multi-cell traps at high magnetic fields [17] based on the "diocotron excitation" of the plasma [18]. This technique does not seem suitable for micro-trapping due to the small radius of the micro-tubes and the small distance between them (e.g. 50μm in radius and 200μm apart).

## 1.3 Background

To introduce our approach to the problem, in this section we will review some of the physics of the problem and the tools to study and quantify the results.

Assuming the magnetic field varies slowly in the helical motion length scale, the motion of the particles is *adiabatic*. This can be written as [19],

$$\tau_{cyc}\frac{dln(B)}{dt} \ll 1 \qquad (2)$$

where $\tau_{cyc}$ is the gyro-period of the particle in the magnetic field $B$. A cylindrically symmetric magnetic field and small variation of the field in the transverse plane where the beam positioned is assumed. *Theorem of adiabatic invariance* states the magnetic flux encircled by the particle trajectories remains a constant of motion [20]. In the case of adiabatic evolution the action integral is a constant of motion. Thus,

$$J_i = \oint p_i dq_i = const, \qquad (3)$$

in which $J_i$ is the $ith$ component of the action $J$. If the magnetic field is axially symmetric, this periodic component is $J_\varphi$ in our cylindrical coordinates $(\rho, \varphi, z)$. The $\varphi$ component of the canonical linear momentum is

$$p_\varphi = \gamma m \rho^2 \dot{\varphi} + q\rho A_\varphi, \qquad (4)$$

where $\gamma = \sqrt{1-(v/c)^2}$ with $v$ as the velocity of the particle and $c$ the speed of light. $m$ and $q$ are the mass and the charge of the particle and $A$ is the magnetic vector potential. As a result, using equations (3) and (4) we have,

$$J_\varphi = \int_0^{2\pi} d\varphi\left(\gamma m \rho^2 \dot{\varphi} + q\rho A_\varphi\right) = const. \qquad (5)$$

Busch's theorem implies

$$\gamma m \rho^2 \dot{\varphi} + \frac{q}{2\pi}\psi = const. \qquad (6)$$

where $\psi$ is the magnetic flux encircled by the particle trajectory. Therefore, if the beam radius stays small with respect to the magnetic field, the axially symmetric magnetic field can be approximated to the first order by $B_z(\rho = 0, z)$. As a result Eq. (6) yields,

$$\dot{\varphi} = -\frac{qB_z(\rho, z)}{\gamma m} \qquad (7)$$

Equations (5) and (7) result in

$$2\pi B_z(\rho, z)\rho^2 - \psi = const, \qquad (8)$$

where $\psi \cong \pi\rho^2 B_z(\rho, z)$. Therefore we obtain

$$\psi = \pi\rho^2 B_z(\rho, z) = const. \tag{9}$$

As an example, adiabatic transportation of a beam from a $10^{-2}T$ guiding magnetic field to $3T$ at the trapping area contracts the beam diameter by a factor of $20$.

The transverse component of the magnetic field during the transition from the low field to the high field, compromises the brightness of the beam and cause mirroring. Quantitatively the criterion for not being mirrored by the magnetic field for a particle moving along the $z$ axis is [21]

$$\left|\frac{v_{z0}}{\sqrt{{v_{\rho 0}}^2 + {v_{\varphi 0}}^2}}\right| > \left(\frac{B}{B0} - 1\right)^{1/2} \tag{10}$$

where $v_0$ is the particle velocity and $B_0$ and $B$ are the lower and the higher magnetic fields respectively.

The approach is to shield the magnetic field such that it facilitates the non-adiabatic transportation of the particles to the trap. A design with a broken cylindrical symmetry allows the particles a quick transition from low to high field. Figure 3 shows the magnetic shield and the field values on a cross-sectional plane. The field at the empty space of the Iris is only about one Gauss.

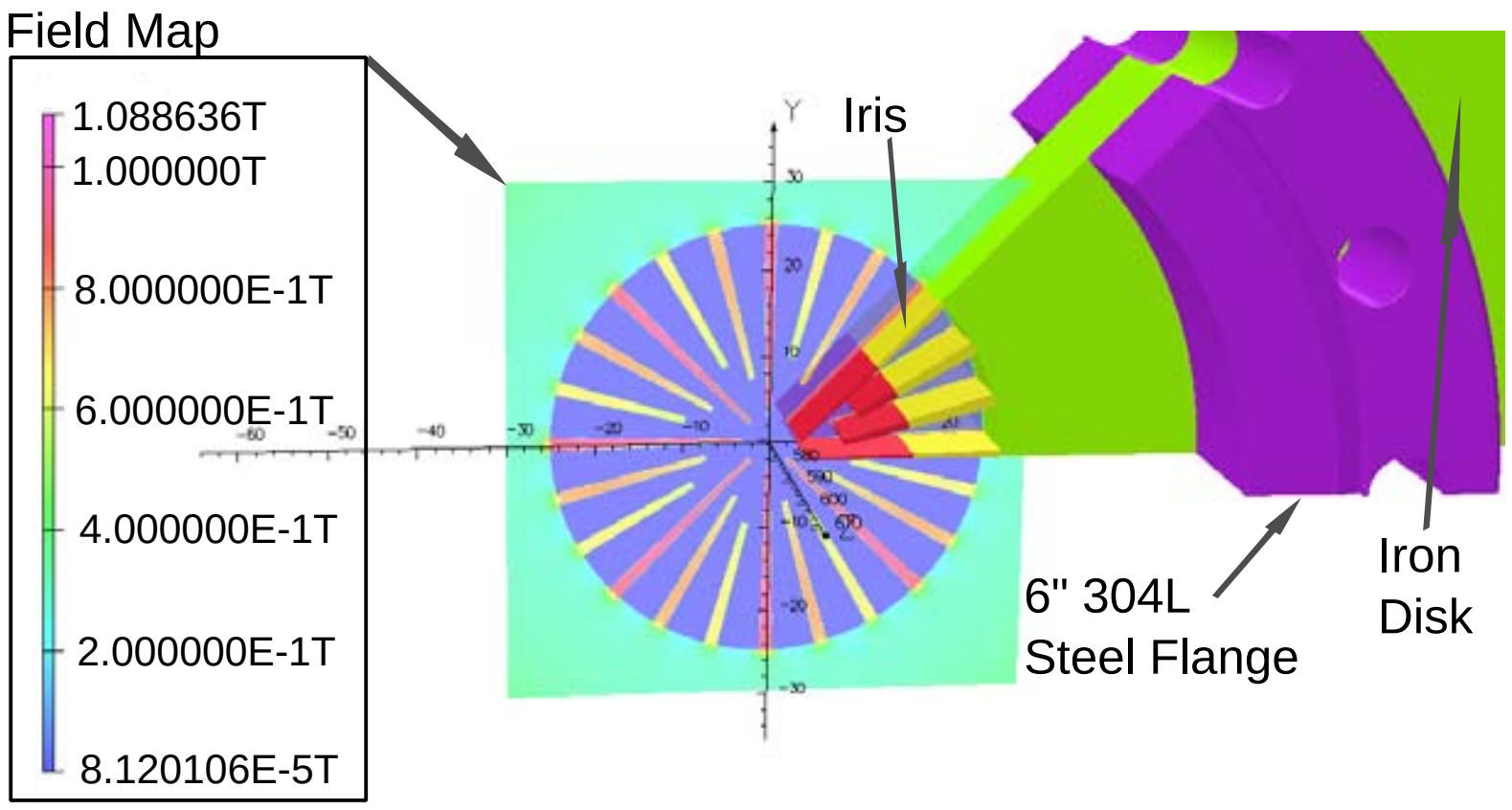

**Figure 3 Magnetic field is evaluated on the plane at the center of the magnetic shield. Only one-eighth of the shield and the flange is presented due the eight-fold cylindrical symmetry. Magnetic field at the empty space of the Iris is only of the order of a Gauss.**

The magnetic shield (traditionally named *Spider*) has been previously employed at Lawrence Livermore National Laboratory [22] to non-adiabatically transport a slow positron from high to low field ($4 \times 10^{-3}T$ to $3 \times 10^{-5}T$). However, as mentioned, for the purpose of Micro-Trapping, the positrons need to be transported from $10^{-2}T$ to $3T$ with minimum beam contraction. The non-adiabatic beam transport in our model takes place from $10^{-2}T$ to $0.3T$ prior to adiabatically travelling to $3T$.

## 2 Method and Simulation

The non-adiabatic low to high field transition of a positron beam is simulated. A design consisting of an *Iris* look opening with 24 narrow equally spaced *Spokes* (Figure 4(a)), a magnetically conducting vacuum sealing midsection (Figure 4(b)), and a disk of radius $70cm$ and $2cm$ thick is employed for magnetic field shielding. The assembled parts are presented in Figure 4(c). The thickness of the *Iris* spokes vary from $1.5mm$ to $0.35mm$. Three different lengths of $22mm$, $17.5mm$ and $15mm$ are employed for the spokes to leave more open area and effectively absorb field lines at higher radii. 65% of the *Iris* is open area. The sharp corners of the *Iris* are not rounded to facilitate the machining process.

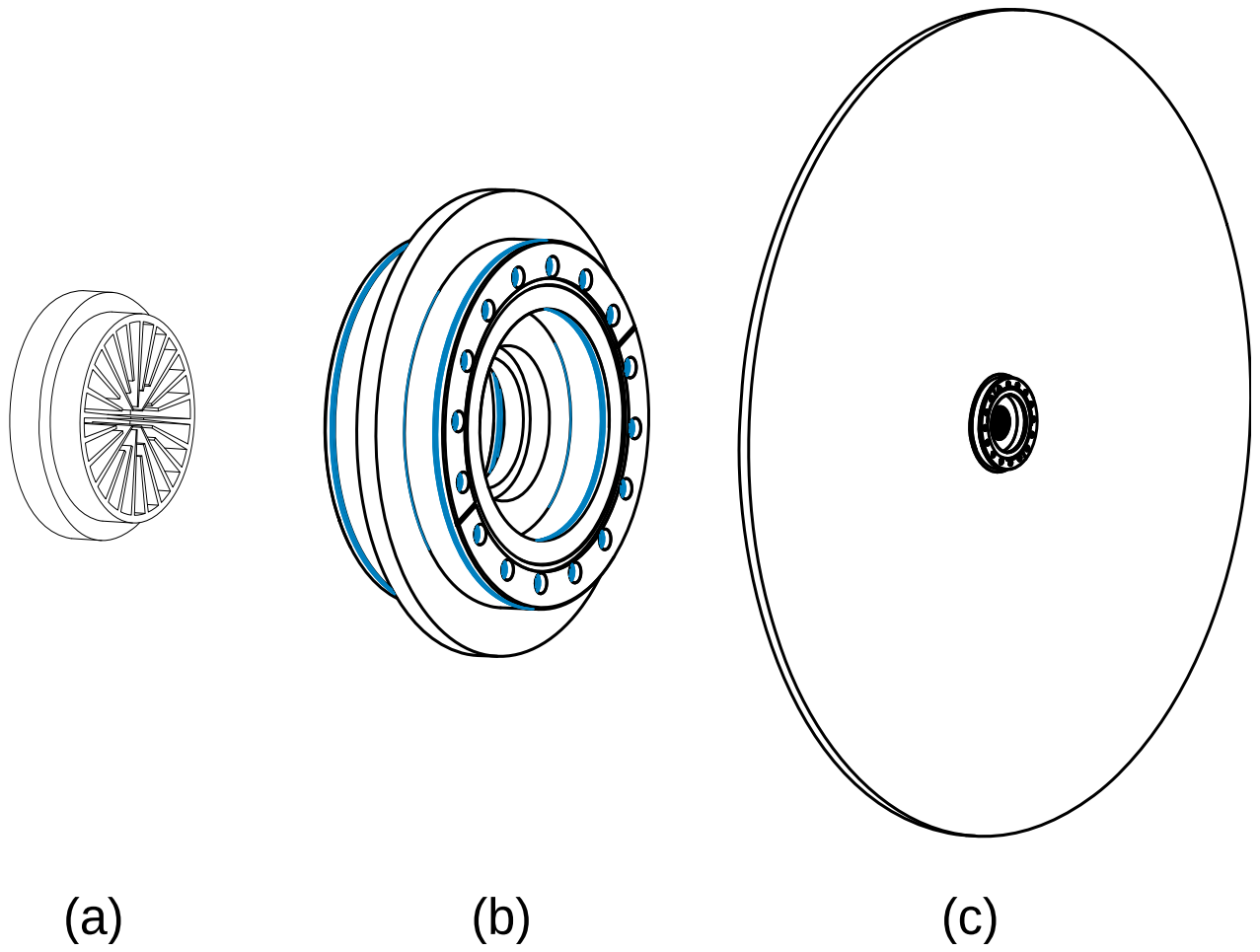

**Figure 4 The different parts of the magnetic shield for low to high magnetic field transportation of slow positrons. (a) The Iris which is located at the center of structure. (b) The middle section of the shield which consists of two flanges for vacuum sealing and a highly magnetic material for good magnetic conduction. (c) The assembly of the three pieces including the Iris, the middle section, and the disk.**

The material used for both the design is Armco Pure Iron [23] with high permeability, low coercivity, and high saturation flux density ($1.4T$). The B-H curve of the material is presented in Figure 5.

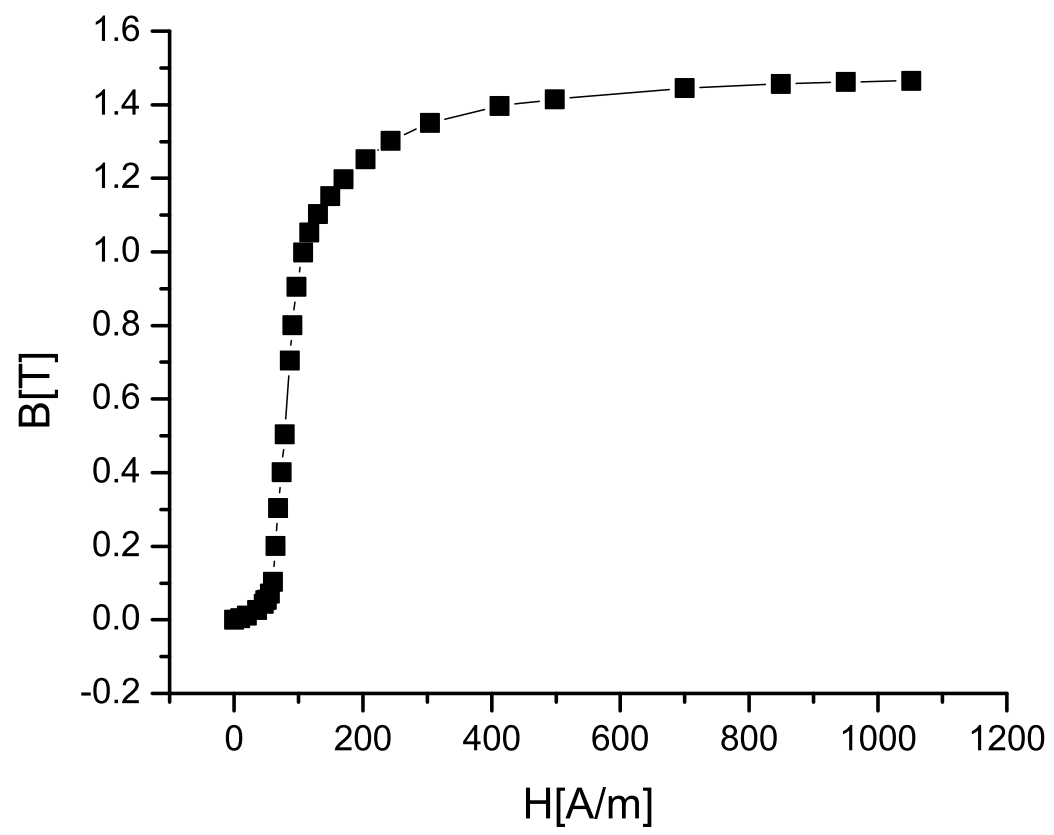

**Figure 5 BH curve of Armco-Smith-Dorf iron is presented.**

The source of the magnetic field is a superconductive $7T\ Magnex\ Model\#7T/160AS$ energized to $3T$. The magnet utilizes active shielding. The main coils of the internal structure are inquired from the manufacturer and imported into the simulation space. The schematics of the model is presented in Figure 6.

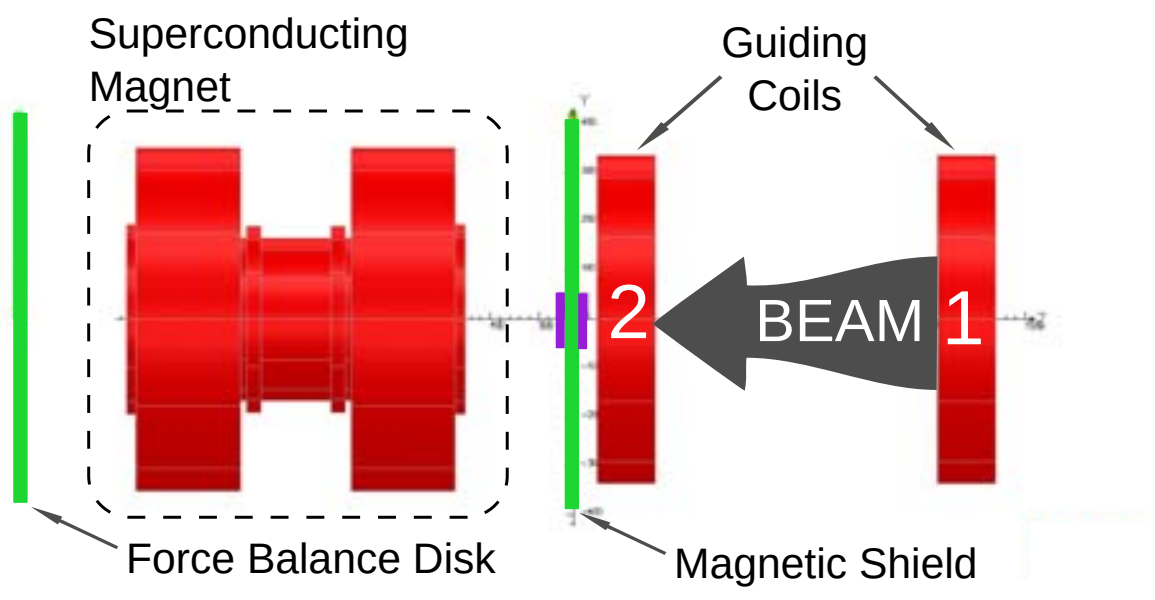

**Figure 6 A schematic of the model is presented. The particles are flown from the center of the first guiding coil on towards magnet. The magnetic shield is presented in green next to the second guiding coil. The second guiding coil is employed to keep the beam radially confined. The purple rectangles (cylinders in 3D) on both sides of the second guiding coil are 6" flanges shown in Figure 4(b) to seal the vacuum.**

A FEA analysis was carried out using the commercially available software OPERA v15.3 [24]. Mesh sizes have been efficiently chosen such that the numerical fluctuations drop below 1% of the actual field values. Eight fold symmetry has been used due to the *Iris* structure to exploit the symmetry.

After solving for the field, it is exported to SIMION 8.0 software to track the particles. The fields of different areas are imported into SIMION using different resolutions to save memory and simulation time. The field around the *Iris* is imported using a grid width of $0.3mm$ as detailed field information is required. Such a resolution requires a considerable amount of computing memory space. For example this small region of the field requires several gigabytes with double precision.

In each run, $5 \times 10^3$ particles were flown for sake of statistics. The beam is investigated on twenty one cross section study planes starting from $1cm$ before to $1cm$ after as well as across the *Iris*. Recalling that the iris itself is $2cm$ thick, each study plane is positioned $2mm$ away from the other.

## 3 Results

Heat treated ARMCO Pure Iron can be employed as an excellent magnetic field absorbent. However, care was taken to prevent a major saturation especially in the area connecting the high filed to the low field . The magnetic field was probed inside and on the surface of the shield for every cubic millimeter, to make sure the criteria were met. Figure 7(a) presents a histogram of the sample field and shows only 0.21% of the points experienced a magnetic field above $1.4T$. Figure 7(b) shows the field contours on the surface of the shield.

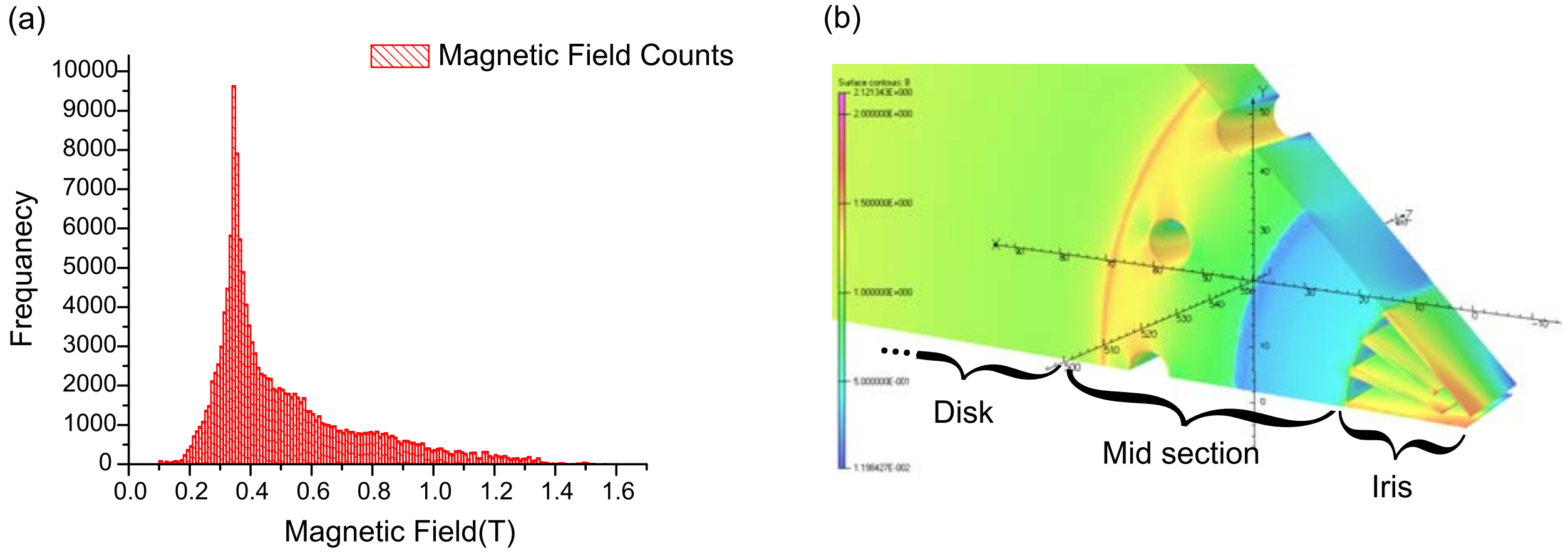

**Figure 7 Magnetic field on the surface of the shield. (a) Frequency count of the magnetic field inside and on the surface of the shield. The field is sampled every cubic millimeter. (b) Magnetic field contours are shown on the surface of the Pure Iron (flange is not shown).**

The magnetic field on the central axis of the magnet are presented in Figure 8 along with the numerical error. The $2cm$ thick disk and the Iris were extended from $z = 560mm$ to $z = 580mm$. The abrupt change in the magnetic field due to the shielding effect is visible in the figure. The field changes by a factor of 40 across the *Iris*.

The magnetic field on the high field side of the *Iris* and at the center of the superconducting magnet are $0.3T$ and $3T$ respectively. Although non-adiabatic transition of the beam across the Iris almost preserves the beam radius, it decreases by a factor of $\sqrt{0.3/3} = 0.32$ due to the adiabatic transition from the Iris to the center of the superconducting magnet.

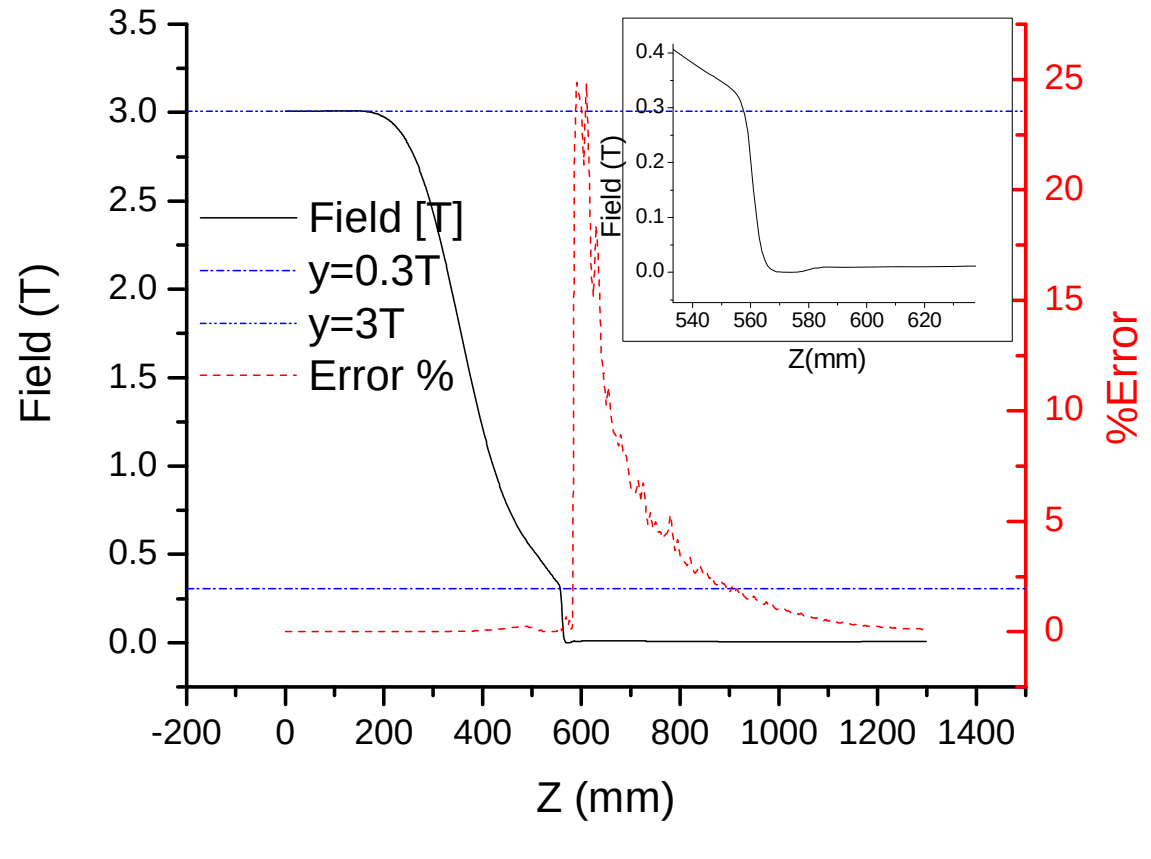

**Figure 8 Magnetic field on axis is presented as well as the FEA numerical error. The field is presented on the left vertical axis in [T], and the error is presented on the right vertical axis in [T]. The inset shows the deviation of the magnetic field in more details just inside the high magnetic field area. The magnetic field increases by a factor of 100 over the length of a one centimeter.**

Adiabaticity of the beam can be evaluated by observing the variations in the beam magnetic field flux using Eq. (9). Since the magnetic flux must be conserved in an adiabatic transportation, changes of this variable indicate a non-adiabatic transport in that region. Figure 9 represents the beam magnetic flux along the beam-line. The largest variation in the flux occurs at the end of the Iris where the field rapidly varies. This sharp field variation is visible in Figure 8 at $Z = 560mm$.

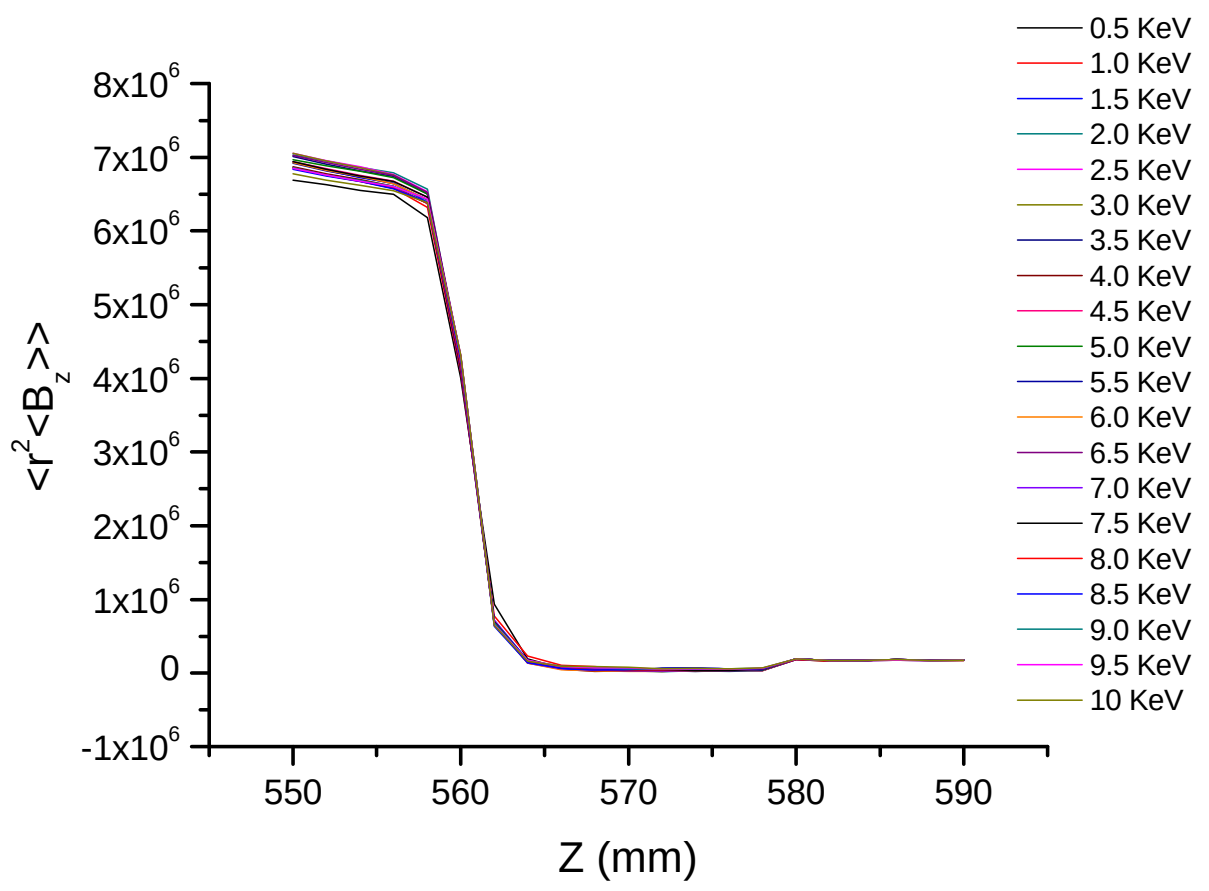

**Figure 9 <r$^2$<B_{z}>> vs Z for different beam energies is presented. The largest change in the beam flux occurs on the higher magnetic field end of the Iris.**

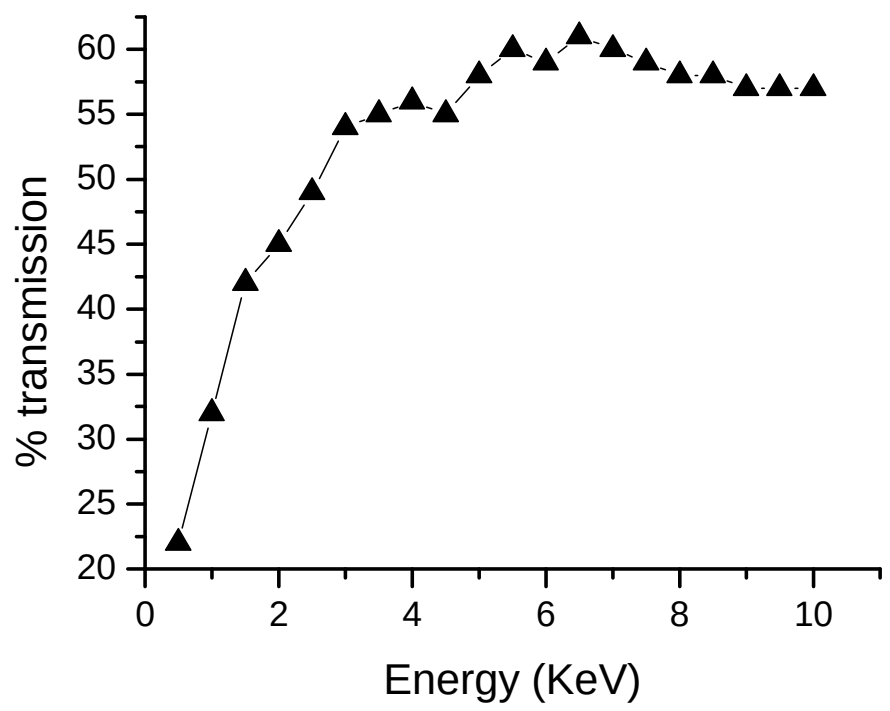

**Figure 10 Transmission efficiency of an initially bright beam at the position of the remoderator. A 61% efficiency is observed for 6.5KeV beam energy.**

To explore the beam transmission effieciency a perfectly bright beam of positrons were flown from the center on the first guiding coil shown in Figure 6 towards the superconducting magnet. The center of the first guiding coil is located $1380mm$ apart from the center of the superconducting magnet. The highest transmission efficiency, approximately $60\%$, is observed for $6.5KeV$ particles, as shown in Fig. 10.

The changes in the magnetic field changed the ratio of the parallel and perpendicular energy components of the beam. Figure 11 represents this ratio $1cm$ away from the Iris in both sides. By comparison with Figure 10, it was seen that the efficiency was the highest when the perpendicular to parallel ratio in the high magnetic field region was minimum. A negative correlation between this ratio and the transmission efficiency is observable.

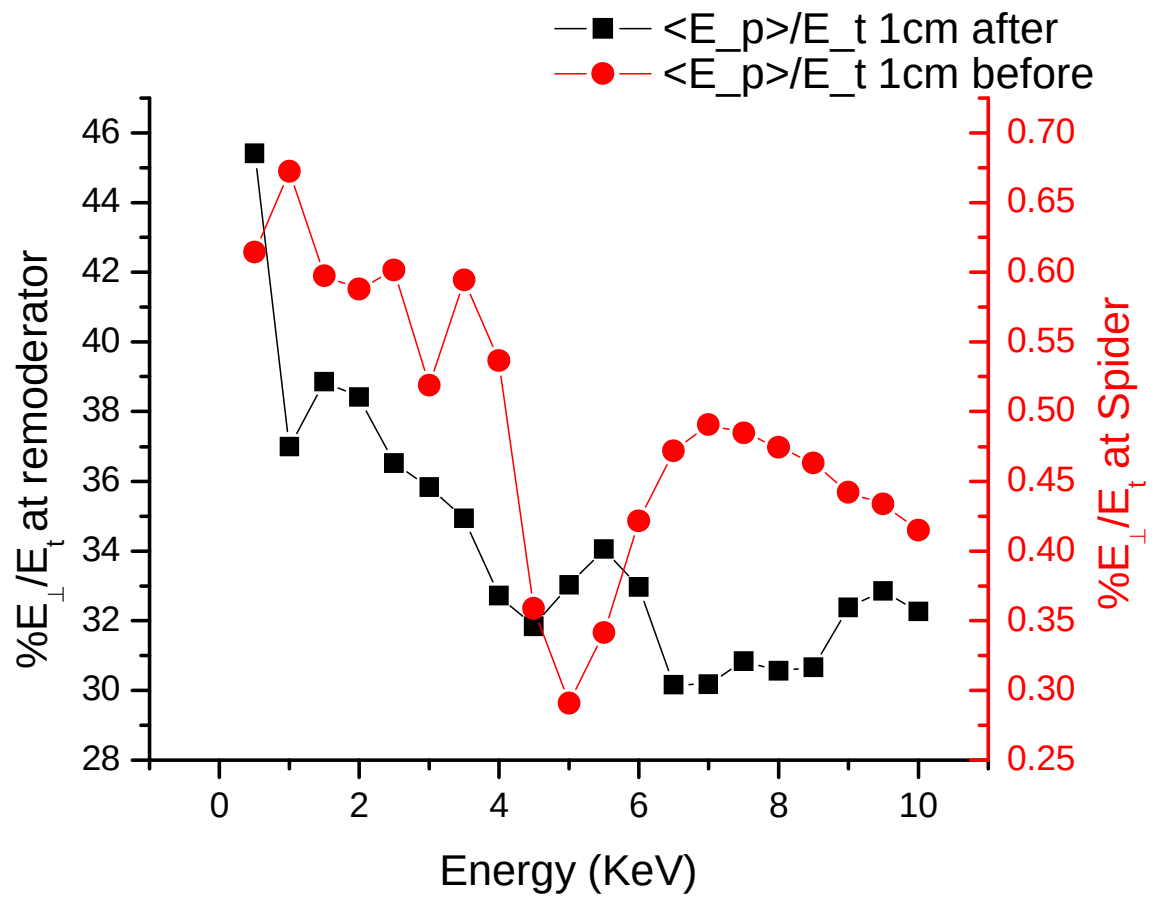

**Figure 11 the ratio of the perpendicular the total kinetic energy of the particles at the position of the remoderator and the spider entrance.**

This indicates that at $6.5KeV$ energy, the beam was the most axially energetic in the high field and more particles could be transferred to the high field.

Figure 12 shows the transmission efficiency of the beam versus the initial beam brightness, initiated at the center of the first guiding coil, through the Iris, at $6.5KeV$. The transmission efficiency is also presented at $1cm$ further into the high magnetic field for comparison.

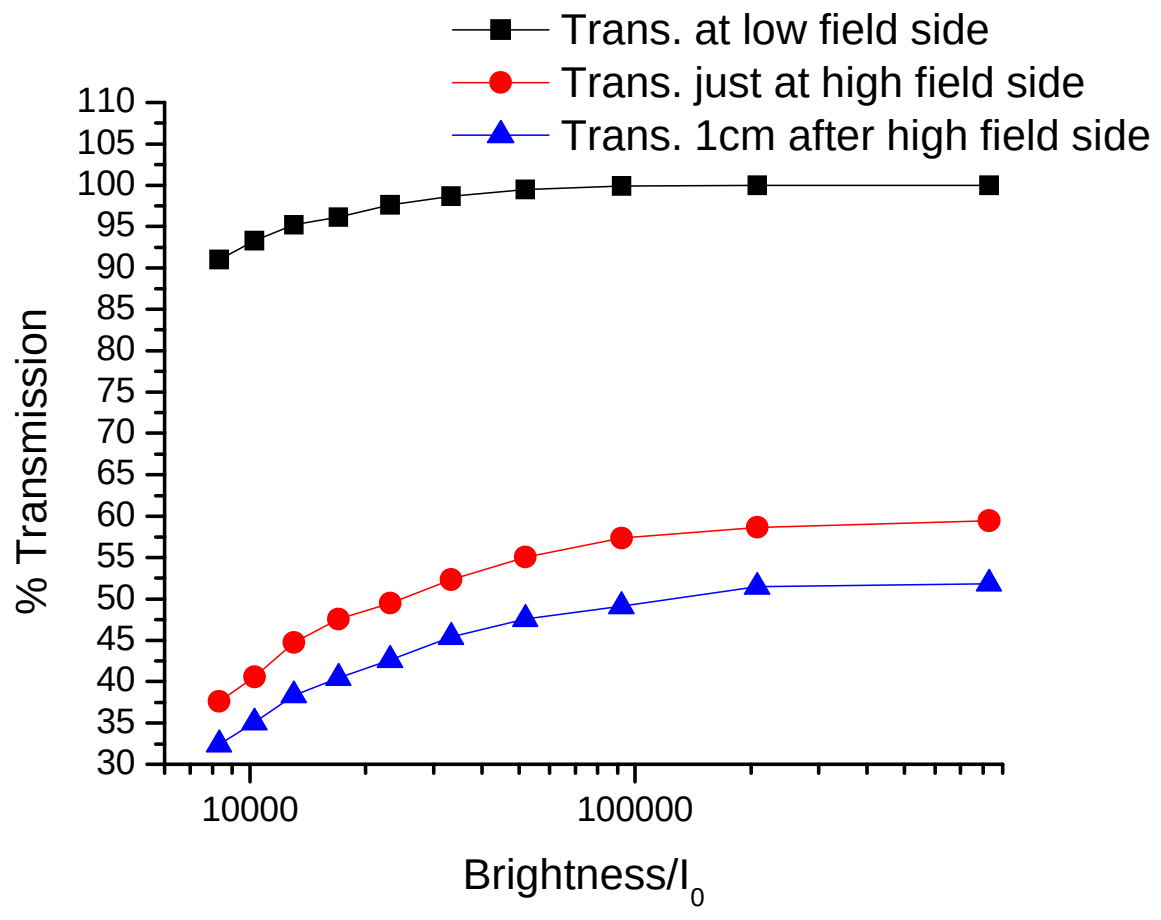

**Figure 12 % beam transmission is presented for a 6.5keV beam for different brightnesses.**

Decrease in brightness compromised the transmission efficiency. Beam brightness can be obtained using [reference needed]

$$B = \frac{d^2 I_0}{dA d\Omega}, \tag{11}$$

in which $I_0$ is the particle flux, $dA$ is the differential illuminated area, and $d\Omega$ is the differential solid angle of the beam.

The efficiency of the transmission did not deviate much at the low magnetic field side of the Iris in comparison with the high magnetic field side.

The distribution of the particles on the cross-section plane of the beam-line is critical for filling the trap micro-tubes. A uniformly distributed beam enables us to fill the micro-tubes in a uniform rate and increase the trapping efficiency. However, the Iris disturbs the magnetic field and the cross section distribution of the beam. A distribution entropy can be defined to quantify the uniformity of the beam. This can be done by dividing the cross section plane into seeds. By counting the number of particles and dividing it by the total number of particles, the particles distribution and entropy can be estimated. We have

$$S = \sum_{seeds} \rho \ln(\rho), \qquad (12)$$

where $\rho$ is the particle distribution on the cross-section plane. $S_0$ can be found using Eq. (12) for a uniformly distributed beam. Figure 13(a) shows the beam after passing through the Iris, and Figure 13(b) represents the cross section of a uniformly distributed beam of the same size with maximum entropy.

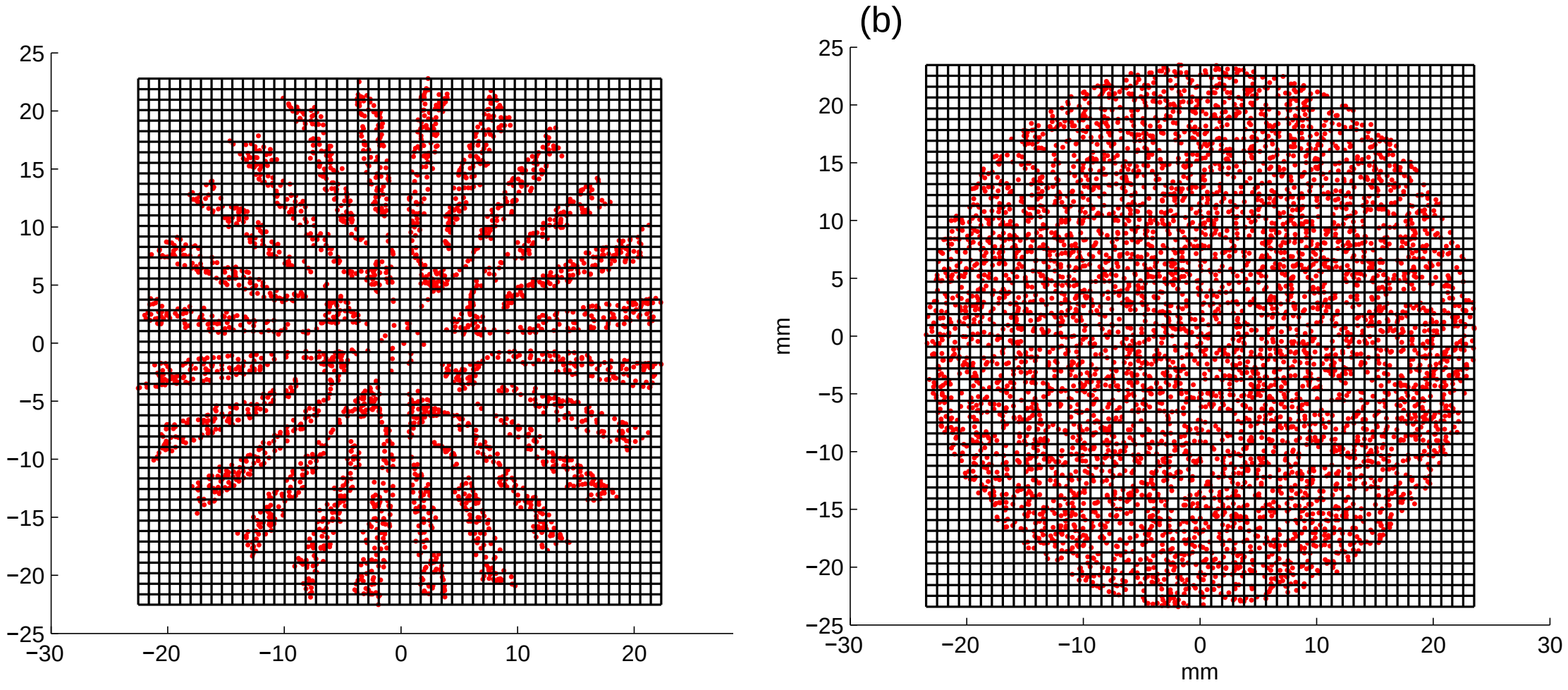

**Figure 13 Beam cross section distribution in $50 \times 50$ grid. (a) the beam cross-section after passing through the Iris. The particles distribution is non-uniform due to the disturbance of the magnetic field. (b) the same number of particles uniformly distributed in the same size beam for comparison. The uniform distribution is used for maximum entropy estimation.**

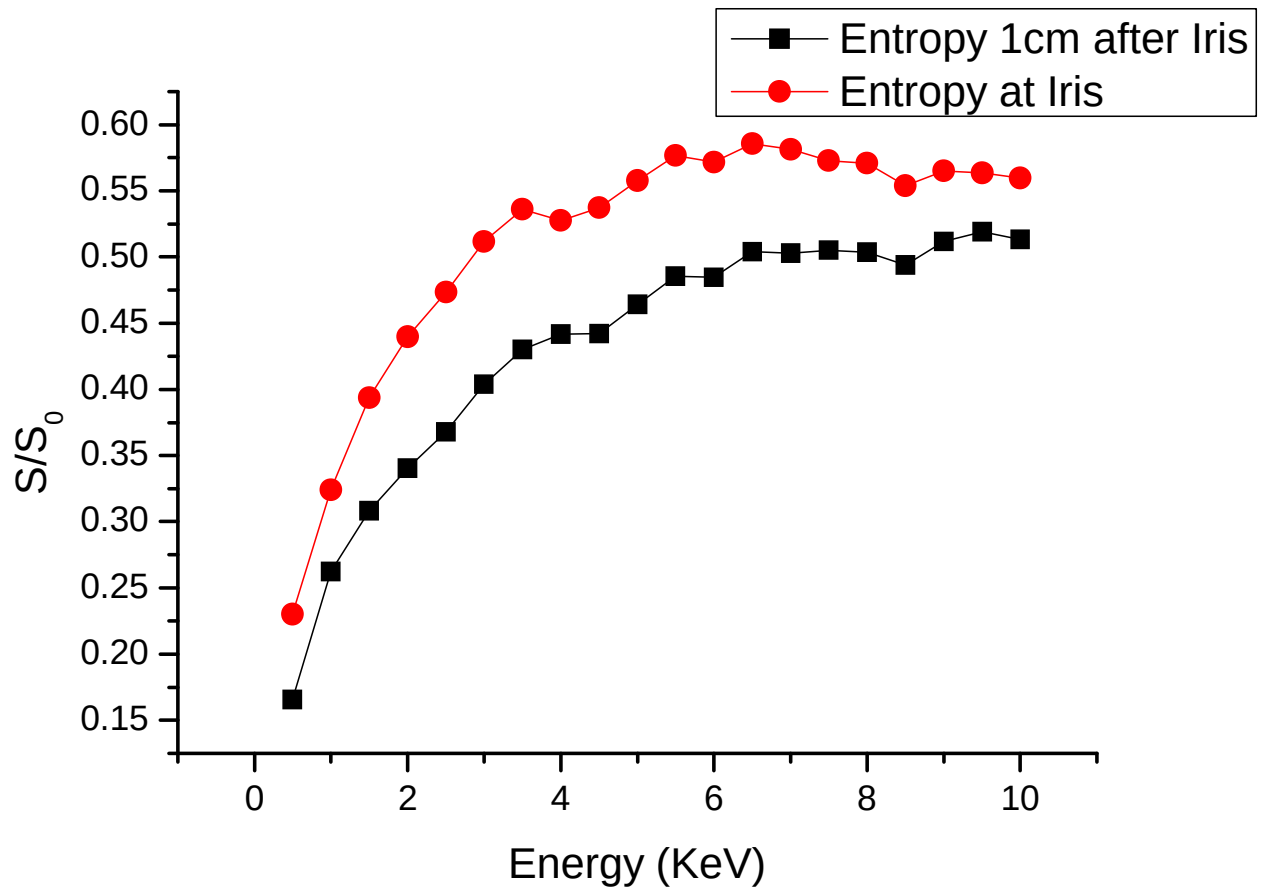

**Figure 14 The normalized entropy of the beam cross-section is presented. $S_0$ is numerically calculated for uniformly distributed circular beam using the same number of particles. The higher the entropy $S$ is the more uniformly the particles are distributed on the cross-section plane.**

The relative entropy for different beam energies is presented in Figure 14. Uniformity decreased as the particles traveled further into the high field. This means that to get a more uniform beam at the center of the high field, re-moderation of the beam should be done as close as possible to the Iris.

# 4 Conclusion

A magnetic shield is designed for low to high field transmission of positrons for trapping in Micro-traps. Simulations show this design allows non-adiabatic transportation of particles from $10^{-2}T$ to $0.3T$. $6.5KeV$ positrons show the maximum transmission efficiency below $10KeV$. Beam brightness and uniformity also enhances at this energy. Remoderation of the beam needs to be applied, for slow positron trapping, as close as possible to the Iris to reduce mirroring.